\documentclass[aip,rsi,reprint,graphicx]{revtex4-1} 
\usepackage{graphicx}
\usepackage{dcolumn}
\usepackage{bm}
\usepackage{natbib}
\usepackage{amsmath}
\begin{document}
\title{Pneumatically actuated and kinematically positioned optical mounts compatible with laser-cooling experiments}
\author{Roger C. Brown}
\affiliation{Joint Quantum Institute, NIST and University of Maryland,
Gaithersburg, Maryland 20899}
\author{S. Olmschenk}
\affiliation{Joint Quantum Institute, NIST and University of Maryland,
Gaithersburg, Maryland 20899}
\affiliation{Department of Physics and Astronomy, Denison University, Granville, Ohio 43023}
\author{Saijun Wu}
\affiliation{Joint Quantum Institute, NIST and University of Maryland,
Gaithersburg, Maryland 20899}
\affiliation{Department of Physics, College of
Science, Swansea University, Swansea, SA2 8PP, United Kingdom}
\author{A. M. Dyckovsky}
\affiliation{Joint Quantum Institute, NIST and University of Maryland,
Gaithersburg, Maryland 20899}
\affiliation{Loudoun County Public Schools Academy of Science, Sterling, Virginia 20164}
\author{R. Wyllie}
\affiliation{Joint Quantum Institute, NIST and University of Maryland,
Gaithersburg, Maryland 20899}
\author{J. V. Porto}
\affiliation{Joint Quantum Institute, NIST and University of Maryland,
Gaithersburg, Maryland 20899}

\date{\today}

\begin{abstract}
We present two complementary designs of pneumatically actuated and kinematically positioned optics mounts:  one designed for vertical mounting and translation, the other designed for horizontal mounting and translation. The design and measured stability make these mounts well-suited to experiments with laser-cooled atoms.
\end{abstract}

\maketitle
We describe a pneumatically actuated mirror system with long term repeatable sub-milliradian pointing stability ensured by a 3-point kinematic positioning system.  This system is easily capable of moving 50~mm~(2") optical components nearly the full throw of a pneumatic piston cylinder in $< 1$~s~($\approx$ 36~mm to 50~mm for the designs presented, though longer actuation distances could be achieved with straight-forward modifications).  The vertically oriented design occupies the same space on an optical table as a typical 38~mm (1.5") post, is compatible with fork-type optical clamps, and its actuation requires no additional space on the optical table.
The system components are relatively inexpensive and cost $ < \$ $1200 for a set of 4 vertical and 2 horizontal units.

The first step in the production of atomic quantum gases is typically laser cooling in a 6-beam magneto-optical trap~(MOT)~\cite{firstMOT} loaded from either a 2-D MOT~\cite{2Dplus3DMOT} or a Zeeman-slowed atomic beam~\cite{firstSlower,YUJUPhysRevA2009,BellRSI}.  To create a capture volume sufficient for subsequent evaporation, the optical beams of the MOT are typically 25~mm to 38~mm (1-1.5") in diameter and represent a significant fraction of a vacuum chamber's surface area. However, increasingly complicated quantum degenerate gas experiments demand further optical access for e.g. dipole trapping~\cite{GrimmDipole}~(where translating pneumatic mirrors have previously been employed~\cite{FirstPMirrorinQDG}), optical lattices~\cite{RevModPhys.80.885}, 
mixtures/photoassociation beams~\cite{Ospelkaus2008}, etc.
It may be possible to combine beams using polarizing beam splitters and/or wavelength dependent filters. However, these stationary solutions necessarily constrain the acceptable wavelengths and polarizations of the beams, and therefore can prohibit different classes of experiments.
Another common solution is to use commercially available kinematic ``flipper" mounts~\cite{YUJUPhysRevA2009} which use an electrical motor to rotate optics in and out of a beam path.
These mounts require extra space for the retracted mirror on the optical table.
In addition, mirrors required to accommodate large optical beams can exceed the manufacturer's weight specifications, creating failure modes that occur on several month time scales.
The system of pneumatically actuated optical mounts detailed here addresses these issues, enabling a reliable, inexpensive system with long-term stability.


\begin{figure}
\begin{center}
\includegraphics[width=3in]{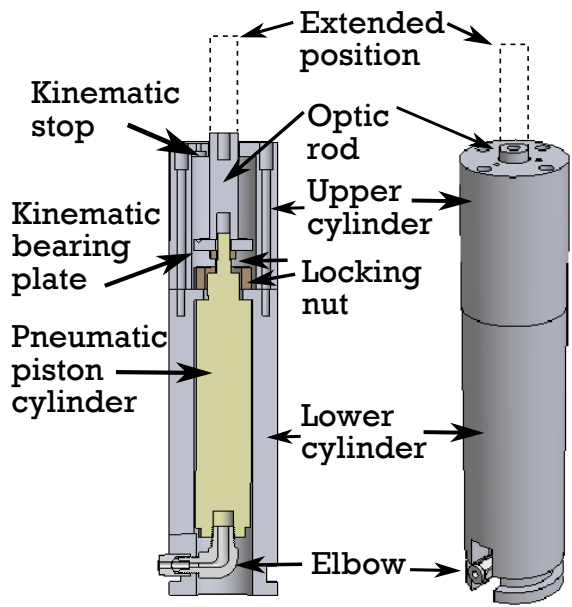}
\end{center}
\renewcommand{\baselinestretch}{1}
\small\normalsize
\begin{quote}
\caption{Cross section and assembled view of vertically oriented and translating optic mount.  The 12.5 mm (0.5") diameter optic rod and piston connected to it below are shown in the compressed position.  The post can be extended the to the full throw of the piston minus a small amount of translation to ensure that kinematic contact is made. The piston throw is 38~mm (1.5") and the throw of the actuated post is 36~mm (1.42").  In the extended position, the kinematic bearings are coincident with the kinematic stops. 
The upper cylinder, lower cylinder, kinematic bearing plate, and optic rod are custom parts. The pneumatic piston cylinder, elbow, and locking nuts are standard commercial products. \label{fig:pneumaticmirrorsectionandside}}
\end{quote}
\end{figure}
The vertically-oriented design, shown in FIG.~\ref{fig:pneumaticmirrorsectionandside}, is based on a single-acting spring return pneumatic cylinder with a 1.5" throw~(SMC Corporation NCMKB088-150CS)\cite{PartNumberNote}.
The overall height of the assembly and choice of piston throw are determined by the height of the beam above the optical table~(which in our case is 25.4~cm (10")) and by the size of the optic to be actuated~(50~mm (2") elliptical gold mirrors), respectively.  The assembly includes 4 custom designed parts: a lower cylinder, an upper cylinder, a kinematic bearing plate, and an optic rod.
The piston is housed in the lower cylinder of the assembly which contains a center hole for mounting the piston, four threaded holes to mount to the upper cylinder and a flange around the bottom compatible with standard fork optical clamps.  The piston is secured to the lower cylinder by a locking nut included with the piston.  The piston actuates a 12.5~mm (0.5") diameter optic rod threaded for mounting to optics and a kinematic bearing plate with three ball bearings to create a repeatable stable mechanical structure~\cite{Maxwell1890}.
The kinematic bearing plate is fixed in place between a locking nut (bottom) and the optic rod (top).
When extended the three ball bearings of the kinematic bearing plate make contact with the kinematic stop hardware screwed into the upper cylinder. We use standard three point kinematic stops with three distinct pieces: a cone, a flat, and a V-cut groove~(Hitek Hardware; KC-1032-TH, KF-1032-TH, and KS-1032-TH).

\begin{figure}
\begin{center}
\includegraphics[width=3.3in]{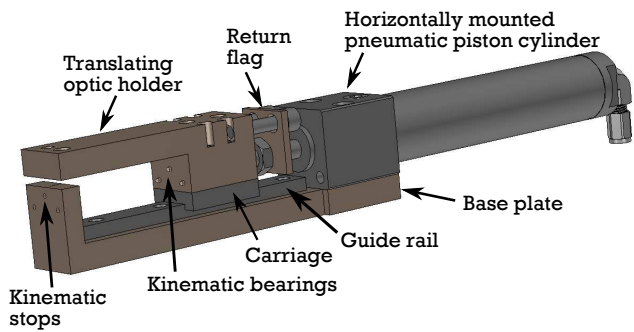}
\end{center}
\renewcommand{\baselinestretch}{1}
\small\normalsize
\begin{quote}
\caption{Horizontally oriented and translating design.  The carriage and piston are shown in the compressed position. In the extended position the kinematic bearings are coincident with the kinematic stops and the carriage assembly is repeatability positioned. The carriage can be translated by nearly the full throw of the piston $\approx$ 50~mm (2").  The base plate, translating optic holder, and return flag are Custom designed. The carriage, guide rail, and pneumatic piston cylinder are commercially available.  \label{fig:Horizontalpneumaticmirrorsystem}}
\end{quote}
\end{figure}

The horizontally-oriented design is based on a horizontally-mounted pneumatic cylinder~(SMC NCMR106-0200CS) and a commercially available linear guide rail and carriage system~(IKO Nippon Thompson; LWLF14R150BPS2 and LWLF14C1BPS2).  The design includes three custom pieces shown in FIG.~\ref{fig:Horizontalpneumaticmirrorsystem}. The first is a monolithic base plate containing threaded mounting holes for the horizontally oriented piston, pedestal mounts, and linear guide rail as well as a three counter bores to hold the kinematic hardware as discussed in vertical design.  Next, the translating optic holder is designed to be mounted on the moving carriage and has threaded holes for the specific optic mount to be translated.
It also contains three press-fit ball bearings to make the necessary 3-point contact with the kinematic stops.  The end of the piston, padded by a rubber stopper, presses on this piece in the extended position.  The final custom piece is a return flag which allows the carriage to be retracted by the piston.  As the piston retracts, the flag catches on a nut around the piston.

\begin{figure}
\begin{center}
\includegraphics[width=2.7in]{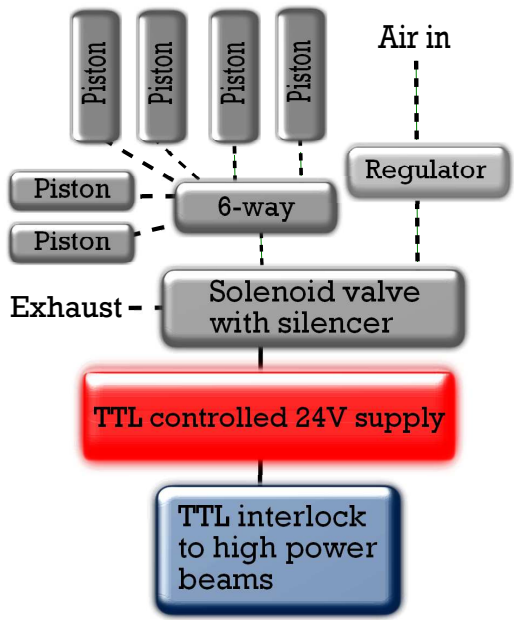}
\end{center}
\renewcommand{\baselinestretch}{1}
\small\normalsize
\begin{quote}
\caption{Block diagram of pneumatic mirror system. Dotted lines indicate pneumatic connections, solid lines indicate electrical connections. \label{fig:pneumaticmirrorsystem}}
\end{quote}
\end{figure}

The pneumatic control system shown in FIG.~\ref{fig:pneumaticmirrorsystem} is used to actuate the mounts.
Laboratory compressed air regulated to 330~kPa~(48~PSI) minimizes the vibrational disturbance of other optics while maintaining a reasonable actuation time.  An electrically controlled solenoid valve~(SMC VQC2101-5 mounted to SMC VQ2000-PW-02T) switches the air to a 6 way splitter~(SMC KQ2ZT01-34S) for all pistons simultaneously. Most exhaust gas is released at the solenoid valve through 35~dBA silencers~(SMC AN202-02) which can be located meters away from the experimental chamber, reducing acoustic noise and air currents as well as electrical noise associated with the switching of the solenoid valve. We use 5/32"~OD
tubing before the splitter~($\approx$~3~m) and 1/8"~OD tubing between the splitter and the pistons ($\approx$~1~m). The solenoid valve is TTL controlled so that pistons can be synchronized with the rest of the experimental cycle.  It is important to consider interlocking the position of the piston with respect to the on/off state of the other beams behind it. This can prevent safety hazards associated with unintentionally scattered laser light or, in the case of higher powered beams, damage to the piston and optics on it.  In our experiment we accomplish this by interlocking the piston TTL signal to the high-power dipole trapping beam TTL signal.

\begin{figure}
\begin{center}
\includegraphics[width=3.3in]{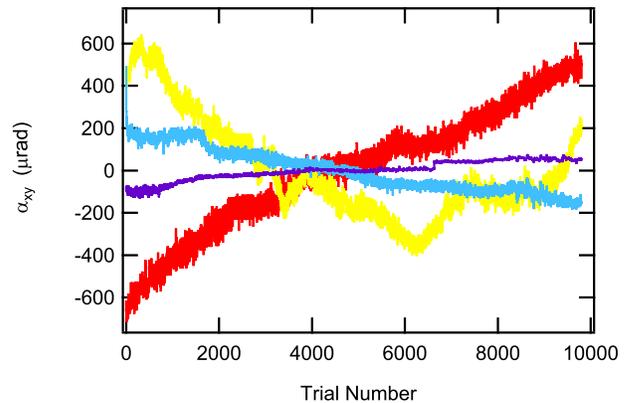}
\end{center}
\renewcommand{\baselinestretch}{1}
\small\normalsize
\begin{quote}
\caption{Pointing stability graph; The points are the x, y angular displacements, $\alpha_{x,y}$, in $\mu$rad (corresponding to spatial displacements $<$~1.5~mm, while the length of a CCD pixel was 6.4~$\mu$m) of the peak centers on the CCD located 165~(1)~cm away from the actuated mirror.  The vertical pneumatic design is in blue along x and violet along y while the horizontal pneumatic design is in red along x and yellow along y.  \label{fig:pointingstabilitygraph1}}
\end{quote}
\end{figure}
We characterize the mechanical stability of the mounts in terms of the position stability of the optical beams they actuate (which for example would lead to intensity/number fluctuations in a MOT). To make position stability measurements path length independent, we measure the angular pointing stability. We  reflect a collimated 635~nm laser beam from a single-mode fiber with a $1/e^2$ beam diameter of 700~$\mu$m off the actuated mirror, and image the position of the beam on a CCD beam profiler~\cite{ISOstandardNote, PointingStabilityNoise}.
In all test measurements the mirror was an elliptical 25~mm~(1") diameter gold mirror epoxied onto a 25~mm~(1") diameter post angled at 45$^o$ which weighed $\approx45$~g (the 50~mm~(2") elliptical gold mirrors used in the apparatus are $\approx100$~g).  An individual position measurement cycle is 4~s with 1.6~s to actuate the mirror and allow vibrations to damp~(the vibrations on the optical table were measured to damp out in $<$~100ms using an accelerometer) 
, then a 100~ms exposure on the CCD, 300~ms dark time, and another 2~s to retract and stabilize the mirror.
This cycle was repeated 9800 times over $\approx$~12~hrs where the temperature in the room was 21$^o$C and fluctuated by $\leq$~0.5$^o$C.  We estimate the angular pointing stability from the spread of the data points in FIG.~\ref{fig:pointingstabilitygraph1} and the beam path length from the pneumatic mirror to the beam profiler, 165~(1)~cm~\cite{CommercialDataNote} (In the experimental apparatus, 30~cm beam paths are used.).  We also estimate ``short term" pointing stability by binning the data into 100 shot intervals and computing the standard deviation for each interval. Averaging over all collected intervals, the short term standard deviation is less than 35~$\mu$rad while, the standard deviation of the worst 100 shot interval was 300~$\mu$rad.  
As a final more qualitative discussion of repeatability/durability, we note that our system of four vertically oriented cylindrical mirrors has functioned successfully for more than 2~years without replacement of any mount components. We tweak up the alignment of the mirrors every 2 to 4 months.

In summary, we have designed and characterized a system of durable, pneumatically actuated, and kinematically positioned mirror mounts.  The mounts have an average short term standard deviation of less than 35~$\mu$rad and several-thousand-cycle angular reproducibility of better than $\pm$1~mrad.

We thank Karl D. Nelson, Martin Zelan, Aaron Young, Silvio Koller, and David Norris for discussions and technical contributions. We thank L. J. LeBlanc and S. Sugawa for critically reading the manuscript. We acknowledge financial support from the ARO Atomtronics MURI.

\begin{thebibliography}{15}
\bibitem [{Par()}]{PartNumberNote}%
  \BibitemOpen
  \href@noop {} {}\bibinfo {note} {All product names and part numbers are for
  specificity only and do not constitute an endorsement by the authors or their
  parent institutions.}\BibitemShut {Stop}%
\end{thebibliography}
The total cost for all commercially available components, at the time of construction, for a 4 vertical and 2 horizontal unit system, excluding machining, is shown in table \ref{Table1}.\\
\\
\begin{table}[ht]
\caption{\label{Table1} Parts and price list}
\begin{ruledtabular}
\begin{tabular}{ldd}
Item (quantity) & \multicolumn{1}{c}{\textrm{Unit price ($ \$ $)}} & \multicolumn{1}{c}{\textrm{Total price ($ \$ $)}} \\
\colrule
Pressure regulator & 20 & 20\\
Solenoid valve + silencers & 100 & 100 \\
6 way splitter & 15 & 15\\
Tubing + connectors & 150 & 150 \\
Pistons (6) & 20 & 120\\
Screws + bearings & 20 & 20 \\
Linear rail and carriage (2) & 145 & 290 \\
24V power supply for solenoid & 20 & 20\\
Kinematic hardware(6) & 65 & 390 \\
\colrule
Total &  & 1125 \\
\end{tabular}
\end{ruledtabular}
\end{table}

\end{document}